\documentclass[prl,twocolumn,showpacs,preprintnumbers,amsmath,amssymb]{revtex4-1}
\usepackage{dcolumn}% Align table columns on decimal point
\usepackage[usenames,dvipsnames]{color}

\usepackage{graphicx}
\usepackage{amssymb}
\usepackage{amsmath}
\usepackage{bm}
\usepackage{epsfig}
\usepackage{ulem}

\newcommand {\be}{\begin{equation}}
\newcommand {\ee}{\end{equation}}
\newcommand {\bea}{\begin{eqnarray}}
\newcommand {\eea}{\end{eqnarray}}
\newcommand {\FIG}[1]{Fig. \ref{#1}}
\newcommand {\FIGS}[1]{Figs. \ref{#1}}

\newcommand {\PRE}[1]{{Phys. Rev. E} {\bf {#1}}}

\newcommand {\PRL}[1]{{Phys. Rev. Lett.} {\bf {#1}}}

\newcommand {\SCI}[1]{{Science} {\bf {#1}}}

\begin{document}

\title{Bond-site duality and phase transition nature of explosive percolations on a two-dimensional lattice}
\author{Woosik Choi}
\author{Soon-Hyung Yook}\email{syook@khu.ac.kr}
\author{Yup Kim} %\email{ykim@khu.ac.kr}
\affiliation{Department of Physics and Research Institute for Basic
Sciences, Kyung Hee University, Seoul 130-701, Korea}
\date{\today}
\begin{abstract}
To establish the bond-site duality of explosive percolations in 2 dimension, the site and bond explosive percolation models are carefully defined on a square lattice. By studying the cluster distribution function and the behavior of the second largest cluster, it is shown that the duality in which the transition is discontinuous exists for the pairs of the site model and the corresponding bond model which relatively enhances the intra-bond occupation. In contrast the intra-bond-suppressed models which have no corresponding site models undergo the continuous transition and satisfy the normal scaling ansatz as ordinary percolation. 
\end{abstract}


\maketitle

%\section{Introduction}
%Even though percolation has been applied 
%to many phenomena \cite{Stauffer_book}, it was considered a mature field of physics. 
Recently {\bf Achlioptas process (AP)} \cite{Achlioptas09} which was suggested to show a supposedly first order transition on the complete graph triggered intensive studies on explosive percolations \cite{Radicchi10,Cho09,daCosta10,Lee11,
Riordan11,Grassberger11}.  
However subsequent studies have proved that the transition in AP on the complete graph is continuous \cite{daCosta10,Lee11,Riordan11,Grassberger11}.
We have also shown that AP on the Bethe lattice shows a continuous transition \cite{Chae12}.
Therefore the transition in the original AP 
is physically established to be continuous in the mean-field level or in high dimensions. 

Until now studies on AP have been done mainly on the complete graph. 
Even though there are some studies on AP in 
{\bf 2 dimension (2$d$)} \cite{Ziff09,Radicchi10,YKim10,Choi11,Bastas11,
Grassberger11},
the transition nature in lower-dimensions is still not fully understood. 
%bond
For example, the bond percolation under AP with a product
rule  was first argued to show a discontinuous transition
\cite{Radicchi10,Ziff09}.
However, based on
the measurement of the largest cluster distribution, Grassberger {\it et al.} \cite{Grassberger11}
argued that the bond percolation with the same product rule in 2$d$ undergoes continuous transition  \cite{Grassberger11}.
%% site
The site percolation under AP with a product rule in 2$d$ has been first proved to undergo the discontinuous transition based on the detailed analysis of cluster size distribution and hysteresis \cite{Choi11}.
In contrast Bastas {\it et al.} argued that the site percolation under AP with a sum rule in 2$d$ undergoes
continuous transition based on the finite-size scaling analysis
with relatively small system sizes \cite{Bastas11}.

Such controversies \cite{Radicchi10,Ziff09,Grassberger11,Bastas11,Choi11}
also imply that there doesn't seem to exist the bond-site duality among explosive percolation models
on 2$d$ lattices
unlike ordinary percolation \cite{Stauffer_book,Sykes64}.
Here the bond-site duality means that a bond percolation model has
the same transition nature or belongs to the same universality class
as the corresponding site percolation model except for the properties depending
on details of models such as transition probability $p_c$, etc. 
Moreover, if the transition is truly discontinuous, then determination of critical exponents from finite size scaling analysis like in Ref. \cite{Bastas11}
has no physical meaning as we already addressed in Ref. \cite{Choi11}.
Thus, resolving such controversies by establishing the bond-site duality in 2$d$
is theoretically very important and interesting.

The controversies should come from ambiguities in the definitions of explosive percolation models
on lattices. Therefore it is very important to make clear definitions of models with various growth rules for AP. There can be six kinds of models on a 2$d$ square lattice.  
Among them, two pairs are bond models. One pair consists of the bond models which physically enhance occupation of intra-cluster bonds.
The other pair consists of bond models which relatively suppress occupation
of intra-bonds.

In site percolation there cannot be the distinction between inter-sites and intra-sites to a cluster. So there is no ambiguity in the definition of site models as in bond models.  
As we shall see, the intra-bond-enhanced models and the corresponding site models
show the bond-site duality in which the transition is discontinuous. 
In contrast the intra-bond-suppressed models show
the continuous transition. Physically there should be no site model corresponding to such
intra-bond-suppressed models.

% In order to
%achieve this purpose, we investigate the bond and site percolation 
%under AP with a product and sum rule and show that
%transition appear to vary depending on the model in a 2D lattice.

%\section{Models}
%\subsection{Model}
%  
There are two fundamental percolation models on lattices\cite{Stauffer_book}. One is the site percolation model and the other is the bond percolation model.
In the site percolation, there is no new site occupation which does not change the size of clusters.
Under AP, two vacant sites $A$ and $B$ are randomly selected. 
Let $\left\{s_{A_i}\right\}$ (or $\left\{s_{B_j}\right\}$) be the sizes of
$n_A$ (or $n_B$) clusters which would be connected by occupying the site $A$ (or $B$).
In the square lattice $n_A$ (or $n_B$) is at most 4.
In {\bf the Site model with a Product rule (SP model)} the site $A$ is occupied if
$1\times\prod_{i=1}^{n_A} s_{A_i}<1\times\prod_{i=1}^{n_B} s_{B_i}$.
Otherwise, the site $B$ is occupied. 
Similarly, in {\bf the Site model with a Sum rule (SS model)} the site $A$ is occupied if
$(1+\sum_{i=1}^{n_A} s_{A_i})< (1+\sum_{j=1}^{n_B} s_{B_j})$.

 We define four bond percolation models
under AP. First two unoccupied bonds $a$ and $b$ are selected randomly.
If the bond $a$ is an inter-bond, then
it connects two different clusters of sizes $s_{a1}$ and $s_{a2}$. Then under a product rule the product  $\xi_a$ for the bond $a$ is clearly defined as $ \xi_a \equiv s_{a1} \times s_{a2}$ without any ambiguity. If the bond is an intra-bond, it internally connects two sites in the same cluster. Then $\xi_a$ can be defined in two different ways. One is $ \xi_a \equiv s_{a1} \times 1$ ({\bf Bond Product Type 1 model: BP1 model}). The other is $ \xi_a \equiv s_{a1} \times s_{a1}$ ({\bf Bond Product Type 2 model: BP2 model}). The product $\xi_b$ for bond $b$ is similarly defined.
Then occupy bond $a$ 
 if $\xi_a < \xi_b$. Otherwise occupy bond $b$. Therefore two bond product models, BP1 and BP2, come from the ambiguity 
to define the product for the selected intra-bond.
Similarly we can define two kinds of bond models with a sum rule. The sum $\sigma_a$ for the inter-bond $a$ is defined clearly as $ \sigma_a \equiv s_{a1} + s_{a2}$. In contrast $\sigma_a$  for the  intra-bond $a$ can also be defined in two-different ways: $ \sigma_a \equiv s_{a1} + 0$ ({\bf Bond Sum Type 1 model: BS1 model}) or $ \sigma_a \equiv s_{a1} + s_{a1}$ ({\bf Bond Sum Type 2 model: BS2 model}). Then occupy bond $a$ if $\sigma_a < \sigma_b$. 

The physical meaning of models is that type 1 models (BP1 and BS1) relatively 
enhance the intra-bond occupation, whereas type 2 models (BP2 and BS2) suppress 
the intra-bond occupation. 
Thus if there exists the bond-site duality, 
it should be between type 1 bond models and site models. As we shall see, the duality exists for the pair of BP1 and SP models and the pair of BS1 and SS models.
BP2 and BS2 have no corresponding site models for the duality.  

Until now, only the BP2 model has been studied for explosive bond percolation model   
\cite{Radicchi10,Ziff09,Grassberger11}. In Ref. \cite{Ziff09} BP2 model was argued to
show discontinuous transition, whereas in Ref. \cite{Grassberger11} the same model was argued 
to undergo continuous transition. BP1, BS1 and BS2 models have never been studied until now. Both SP and SS have been studied as explosive site percolation models.
SP model \cite{Choi11} has been proved to show discontinuous transition, whereas SS model \cite{Bastas11}
was argued to show continuous transition based on the numerical studies of relatively small systems.

%%%%%%%%%%%%%%%%%%%%%%%%%%%%%%%%%%%%%%%%%%%%%%%%%%
%  n_s distribution                              %
%%%%%%%%%%%%%%%%%%%%%%%%%%%%%%%%%%%%%%%%%%%%%%%%%%
%\section{cluster size distribution}
 
\begin{figure}[ht]
%\vspace{-0.5cm}
%\hspace{-0.88cm}
\includegraphics[width=8.0cm]{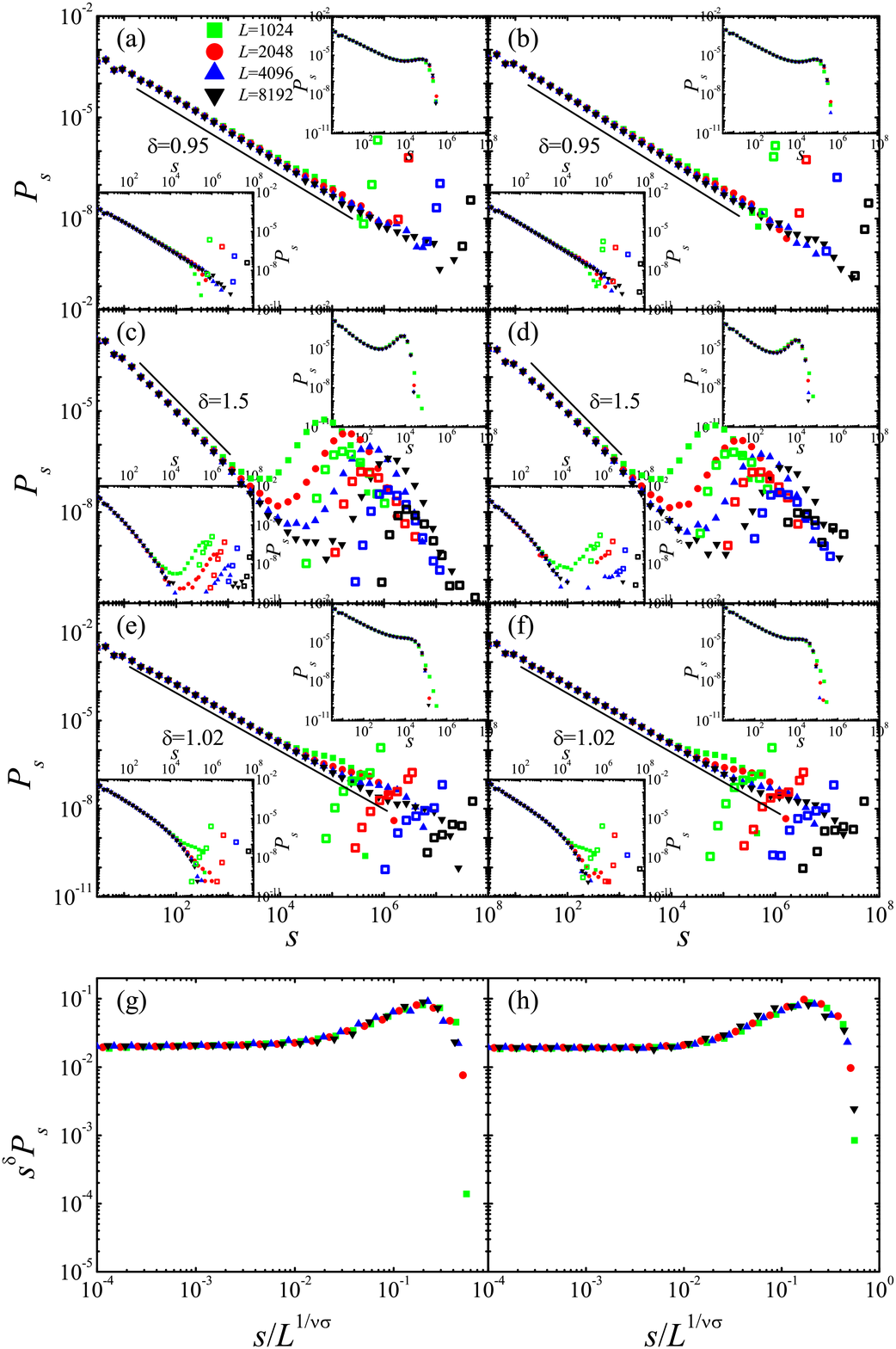}
%\vspace{-1.2cm}
\caption{(Color online) Plots of $P_s(p)$ against $s$ for various models. Unfilled symbols denotes $P_s$ for the largest cluster on the lattice with the linear size $L$.
(a) SP model at $p=0.7723(7)$ ($\approx p_c$). Insets:
at $p=0.75$ $(<p_c)$ (upper inset) and $p=0.78$ ($>p_c$).  The line with $\delta=0.95$
means the relation $P_s \sim s^{-0.95}$.
(b) BP1 model at $p=0.6937(8)$ ($\approx p_c$). Insets: at $p=0.67$ $(<p_c)$ (upper inset) and $p=0.70$ ($>p_c$). The line denotes the same thing as in (a). 
(c) BS1 model at $p=0.5979(4)$ ($\approx p_c$). Insets: at $p=0.590$ $(<p_c)$ (upper inset) and $p=0.605$ ($>p_c$). The line denotes the relation $P_s \sim s^{-1.5}$ 
(d) SS model at $p=0.6916(5)$ ($\approx p_c$). Insets: $P_s$ at $p=0.685$ $(<p_c)$ (upper inset) and $p=0.695$ ($>p_c$). The line denotes the same thing as in (c).  
(e) BP2 model at $p=0.5266(1)$ ($\approx p_c$). Insets: at $p=0.5250$ $(<p_c)$ (upper inset) and 
$p=0.5270$ ($>p_c$).
The line means the relation $P_s\sim s^{-1.02}$.
(f) BS2 model at $p=0.5270(1)$ ($\approx p_c$). Insets: at $p=0.5250$ $(<p_c)$ (upper inset) and $p=0.5275$ ($>p_c$). The line denotes the same thing as in (e).   
(g) Scaling collapse for the relation (\ref{scaling})
of BP2 model and (h) the collapse of BS2 model. 
Used exponents in (g) and (h) are $\delta=1.02$ and $\nu \sigma=0.51$.   }
\label{ps}
\end{figure}
To understand the transition nature of percolation physically, the cluster size distribution should be the first one to understand. The cluster
size distribution $P_s (p)$ at an occupation probability $p$ of a bond (or site) is
the probability that an occupied
site belongs to a cluster which has $s$ sites.
It has been shown that $P_s(p)$ provides an excellent method to determine $p_c$ as well as
the transition nature for {\bf ordinary percolation (OP)}\cite{Stauffer_book} and explosive percolations \cite{Lee11,Choi11}.
Thus, we first investigate $P_s(p)$ for each model
to obtain $p_c$ and the transition nature physically.
%$n_{s}(p;L)$ at a bond (site) occupation probability
%$p$ on a lattice with linear dimension $L$ is defined as the number of
%$s$-clusters per lattice site.
When $p<p_c$, $P_s$ for OP decays exponentially 
as $s$ increases.
%in the limit $L\rightarrow\infty$.
This means that the probability to find
a large cluster vanishes exponentially. On the other hand,
when $p>p_c$, there exists the {\bf macroscopically  large cluster (LC)} and $P_s$ for finite $s$ also decays exponentially with a
peak for LC.
At $p=p_c$ it is well known that $P_s$ satisfies a power-law,
%\bea \label{Ps} 
$P_s(p_c) \sim s^{-\delta}$,
%\eea
with $\delta \simeq 1.055$ for OP. 

In contrast, $P_s$ for
SP model in \FIG{ps}(a) is completely different from that for OP as shown in Ref. \cite{Choi11}.
When $p<p_c$, $P_s$ for SP model has a stable hump in the tail as $p \rightarrow p_c^-$ \cite{Lee11,Choi11}.
The behavior of hump for 
SP model have been studied in detail in Ref. \cite{Choi11}.
%Ps for the macroscopically large cluster
%starts to be detached from the continuous %distribution of Ps
%for microscopic clusters. This detachment behavior seems to
%be independent of L(N) as shown in Fig. 1(c).
On the other hand, at $p\simeq p_c$ $P_s$ for LC 
starts to be detached from the continuous distribution of $P_s$ for finite $s$, which satisfies $P_s \sim s^{-\delta}$ with $\delta =0.95(1)(<1)$.
Such power-law behavior for finite $s$ is observed for sufficiently large $p(> p_c)$ (see the lower inset of  \FIG{ps}(a)) \cite{Choi11}.
Based on this typical behavior of $P_s(p)$  we
determined $p_c$ for SP model \cite{Choi11}. 
Since the hump contribution for $p<p_c$ does not depend on the lattice size $L$ and $\delta<1 $
at $p_c$, there should be many stable
large (but still microscopic) clusters before transition, which strongly indicates the
discontinuous transition for SP model \cite{Choi11}.
To determine the $p_c$'s and find the transition nature for other models,
we now analyze $P_s$'s as in Ref. \cite{Choi11}.

In \FIG{ps}(b)
 we display $P_s(p)$ for BP1 model. The data for $p<p_c$ in the
upper inset of \FIG{ps}(b) clearly shows that the hump in $P_s(p<p_c)$ does
not depend on $L$ as for SP model in \FIG{ps}(a).
At $p\simeq p_c$, $P_s$ for LC starts to be detached from the 
continuous distribution of $P_s$ as shown in the main plot of \FIG{ps}(b).
From the best fit of the data for finite $s$, we
obtain $P_s \sim s^\delta$ with $\delta\simeq 0.95(1) (<1)$.
For $p>p_c$ we find that $P_s$ for finite $s$ still satisfies
the power-law with $\delta<1$ (see the lower inset).
These behaviors of $P_s(p)$ for BP1 model exactly coincides with that
for SP model, which is known to undergo discontinuous transition \cite{Choi11}.
The only difference between BP1 model and SP model is in the value of $p_c$: $p_c \simeq 0.6937(8)$ for BP1 model and $p_c \simeq 0.7723(7)$ for SP model.
The coincidence of $P_s$ between BP1 model and SP model physically means 
that there is the bond-site duality under AP with the product rule if 
the bond model enhances the intra-bond occupation. 

\FIGS{ps}(c) and (d) show  $P_s$ for BS1 and SS models. 
By assuming the relation $P_s \sim s^\delta$ only for small $s$
at $p \simeq p_c$,
we obtain $\delta \simeq 1.5$ for BS1 and SS models. However,
as depicted in \FIGS{ps}(c) and (d),  $P_s$ for BS1 and SS models seems to substantially deviates from the power-law due to a main contribution from the hump distribution for large $s$.
Even for $p_c$ at which $P_s$ for LC starts to be detached
from the seemingly power-law like regime, the hump still exists.
The contribution of the hump distribution part to $\sum_s^{L^2} P(s)=p$ for site percolation
(or $\sum_{s=1}^{L^2} P(s)=1$ for bond percolation) even at $p=p_c$ is more than $90 \%$.
%to the sum rule $\sum_{s=1}^{L^2} P(s)=p$ . 
This means that the contribution of the power-law like regime for small clusters to $\sum_s P(s)$ is trivial. 
This result implies the absence of the singular behavior, $P_s(p) \sim s^{-\delta}$ for $s 
\rightarrow \infty$ at $p_c$, which is the intrinsic property of the continuous transition. Moreover in both BS and SS models the hump distribution for $p<p_c$ does not depends on the lattice size $L$ (see the upper insets).
%However, if $p\simeq p_c$ the peak of hump decreases as $L$ increases,
%which indicates that the contribution of the hump is diminished
%in the limit $L\rightarrow\infty$. Indeed, the fraction
%of sites in the hump near $p_c$ is more than $0.9$,
%and does not depend on $L$. 
This anomalous behavior suggests that
there exists a type of powder-keg \cite{Friedman09}
in BS1 and SS models,
unlike in the fully connected networks \cite{Riordan11}.
In addition, the existence of the hump distribution even for
$p>p_c$ (see the lower insets) implies the existence of multiple stable
macroscopic clusters for $p\ge p_c$ in BS1 and SS models.
Furthermore, $P_s$ for BS1 model (\FIG{ps}(c)) is nearly identical to $P_s$ for SS model (\FIG{ps}(d)) as the case for BP1 and SP models. 
The coincidence of $P_s$ between BS1 and SS models except for the value of $p_c$ 
also implies that there is the bond-site duality under AP with the sum rule if 
the bond model enhances the intra-bond occupation.   
%Moreover, in APBSR1 and APSSR models, when $p>p_c$ there still
%exists a large hump in $P_s$ distribution, which also provides
%another evidence to exist multiple stable clusters above the threshold.
%We will discuss about it in detail in the following sections.

%BP2,BS2
In \FIGS{ps}(e) and (f) we display $P_s$ for BP2 and BS2  models.
$P_s$ for BP2 and BS2 models is physically different from $P_s$ for BP1 and BS1 models.
Even though $P_s$ for BP2 and BS2 models has a hump in the tail
when $p<p_c$ (upper inset), this hump vanishes in the limit $L \rightarrow \infty$ and
$P_s$ at $p_c$ shows the power-law singularity
$P_s(p) \sim s^{-\delta}$ with $\delta =1.02(1) >1$.
As we increase $p$ further to $p>p_c$, $P_s$ for finite $s$ decays exponentially as in OP
(see the lower insets for $p>p_c$).
This $P_s$ behavior is physically the same as that in OP. 
As for OP, $P_s$ or $n_s(\equiv P_s/s)$ at $p_c$ 
for the percolation with continuous transition satisfies the scaling relation
\cite{Stauffer_book} 
\be
\label{scaling}
P_s = s^{-\delta} f(s/L^{1/\sigma \nu})~~ (n_s = s^{-\tau} f(s/L^{1/\sigma \nu})),
\ee
where $\delta=\tau-1$ and $\nu$ is the correlation length critical exponent.
%\begin{figure}[ht]
%\vspace{-0.5cm}
%\hspace{-0.88cm}
%\includegraphics[width=8.5cm]{psscaling.eps}
%\vspace{-1.2cm}
%\caption{(Color online) Scaling collapses for the relation (\ref{scaling})
%    of ordinary bond percolation (a), BS2(b) and BP2 (inset of (b)) models. Used exponents in (b) 
%    are $\delta=1.02$ and $\nu \sigma=0.51$.  
%}
%\label{psscaling}
%\end{figure} 
As shown in \FIGS{ps}(g) and (h),
$P_s$(or $n_s$) for both BP2 and BS2 models satisfies the scaling relation (\ref{scaling})
with $\delta=1.02(\tau=2.02)$ and $\nu \sigma = 0.51$ very well.  
$P_s$ in \FIGS{ps} (e), (f), (g) and (h)  strongly suggests that both BS2 and BP2 models
show continuous transition.
Our estimation of $p_c=0.5266(1)$ for BP2 model is nearly the same as
those in Refs.\cite{Ziff09,Grassberger11}, even though the two references
were contradictory to each other in transition nature. 

From the obtained $\tau$ and $\sigma$ from Eq. (\ref{scaling}), one can calculate the critical exponents
$\beta$, $\gamma$ and $\nu$, which must be identical to the values evaluated from the {\bf finite-size scaling (FSS)}
properties of  the average size $S(p,L)$ of the finite clusters and the order parameter $P_\infty(p,L)$ if the transition is truly continuous. From $S(p,L)$ for BS2 model obtained by the numerical simulations with $L=512 - 4096$ and the relation 
$p_{max}(L) = p_c + b L^{-1/\nu}$, where $p_{max}(L)$ is $p$ at which $S(p,L)$ is maximal
\cite{Stauffer_book}, we obtain $\nu=1.00(1)$ and $p_c=0.5270(1)$ for BS2 model. 
%The obtained $p_c$ is the same as $p_c$ obtained %from $P_s$ data in \FIG{ps}(f). 
From the relation for the maximal value of $S(p,L)$, $S_{max} \sim L^{\gamma/\nu}$ \cite{Stauffer_book} and the obtained data for $S$ we obtain $\gamma/\nu=1.90(2)$. Thus $S(p,L)$ for BS2 model satisfies
FSS ansatz \cite{Stauffer_book} 
$S(p,L)=L^{\gamma/\nu}f((p-p_c)L^{1/\nu})$ very well
with $\nu=1.00(1)$ and $p_c=0.5270(1)$.% and %$\gamma/\nu=1.90(2)$ 
%very well.
$P_{\infty}(p,L)$ for BS2 model satisfies the FSS ansatz \cite{Stauffer_book} $P_{\infty}(p,L)=L^{-\beta/\nu}g((p-p_c)L^{1/\nu})$
very well with $\beta/\nu=0.04(1)$.
Since $\gamma$ and $\beta$ are related to $\tau$ and $\sigma$ as $\gamma=(3-\tau)/\sigma$ and  $\beta=(\tau-2)/\sigma$ \cite{Stauffer_book}, 
$\nu$, $\gamma$ and $\beta$ \cite{Stauffer_book} obtained from FSS ansatz are consistent with   $\tau$ and $\sigma$ obtained from Eq. (\ref{scaling}) for BS2 model. The obtained exponents are $\tau=2.02(1), \sigma=0.51(1), \nu=1.00(1), \gamma=1.90(2)$ and $\beta=0.04(1)$.  
For BP2 model we obtain identical exponents and scaling relations to those for BS2 model. 
%Exponents for BS2 and BP2 models are summarized as
%\bea
%\label{expo}
%\tau=2.02(1), \sigma=0.51(1), \nu=1.00(1), %\gamma=1.90(2), \beta=0.04(1).
%\eea

% %%%%%%%%%%%%%%%%%%%%%%%%%%%%%%%%%%%%%%%%%%%%%%%%%%%%%%%%%%%%%%
%   2nd largest cluster
% %%%%%%%%%%%%%%%%%%%%%%%%%%%%%%%%%%%%%%%%%%%%%%%%%%%%%%%%%%%%%%
%\section{Behavior of the largest cluster and the %second largest cluster}
\begin{figure}[ht]
\includegraphics[width=8.5cm]{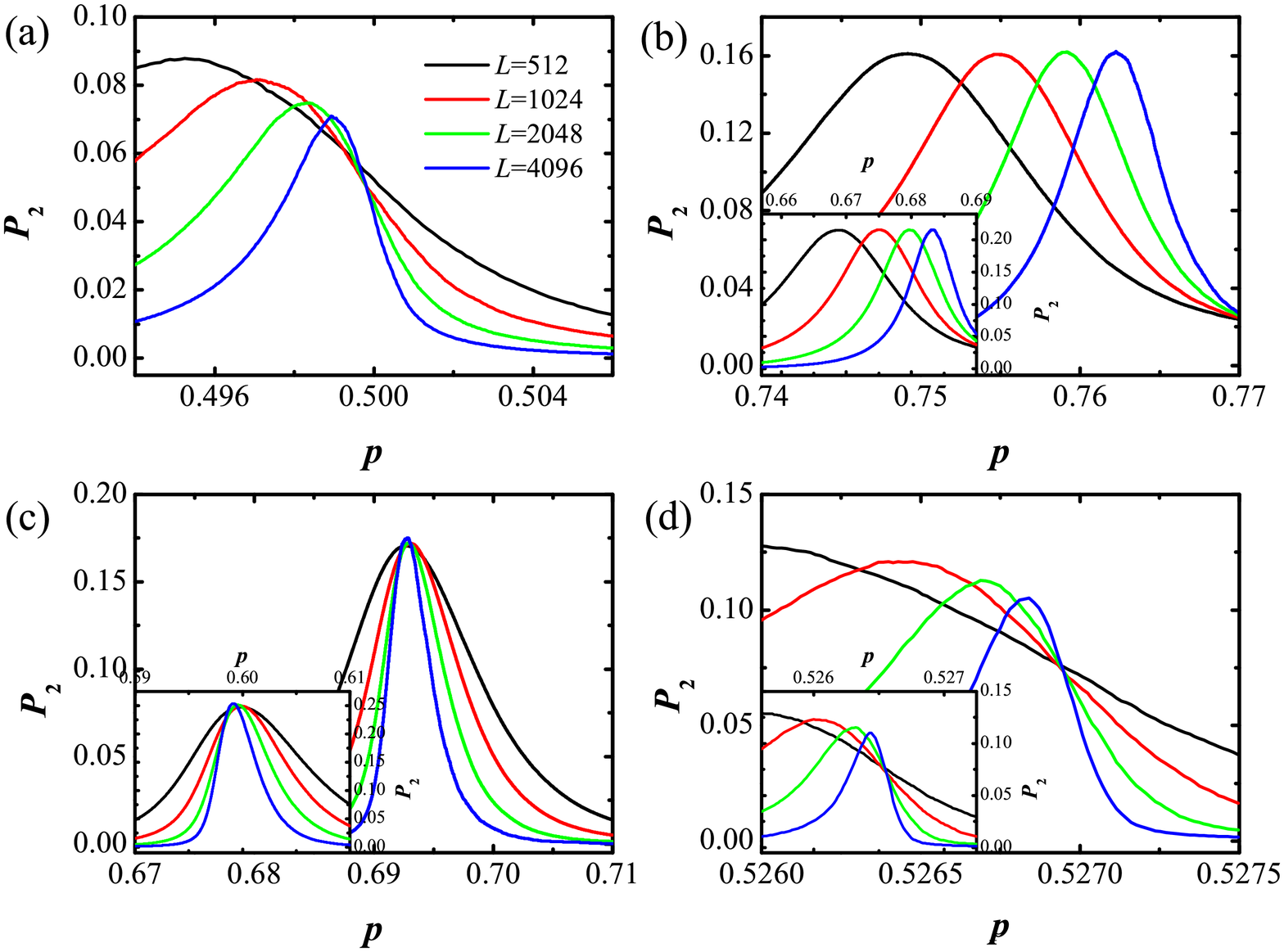}
\vspace{-0.5cm}
\caption{(Color online) (a) Plot of $P_2(p,L)$ against $p$ (a) for ordinary bond percolation, (b) for SP model(Inset:BP1 model), (c) for SS model (Inset:BS1 model) and (d) for BS2 model (Inset: BP2 model).}
\label{p2}
\end{figure}
The discontinuous transition is characterized by the existence of metastable states. The metastable states in percolation transition are generally originated from the coexistence of multiple
macroscopic clusters. 
To study multiple
macroscopic clusters, we now focus $P_2(p,L)$ with which a randomly chosen occupied site belongs to the second largest cluster.
%One is the probability $P_{LC}(p,L)$ with which a randomly selected occupied site belongs to the largest cluster at $p$ on the lattice with the linear size $L$. In the limit
%$L \rightarrow \infty$ $P_{LC}(p,L)$ becomes the order parameter $P_{\infty}(p)$ of percolation transition \cite{Stauffer_book}. The other is 
If the transition is discontinuous, then 
$P_2$ approaches a nonzero value at $p \simeq p_c$ in the limit $L\rightarrow\infty$.
On the other hand, when the transition is continuous,
the largest cluster grows by gradual adding
of small clusters and $P_2$ at $p \simeq p_c$ should decrease to zero in the thermodynamic limit. 
For the sake of comparison, we first measure 
$P_2$ for {\bf ordinary bond percolation (OBP)}.
As shown in \FIG{p2}(a), $p$ at which $P_2$ is maximal, $p_{2max}$, approaches to the known value of $p_c=1/2$ as $L$ increases. Furthermore the maximal value $ P_{2}(p_{2max})$ decreases as $L$ increases. 
These behaviors of $P_2(p,L)$ clearly show the absence
of the multiple macroscopic clusters in OBP for $p \simeq p_c$ in the limit $L \rightarrow \infty$
and the transition becomes continuous.
%\begin{figure}[ht]
%\includegraphics[width=8.5cm]{p2max.eps}
%\vspace{-0.5cm}
%\caption{(Color online) (a) Plot of $P_2(p_{2max})$ against $L$  for SP model. Inset: Plot of $p_{2max}$ against $1/L$.  (b) The same plot for SS model. (c) The same plot for BS2 model. }
%\label{p2max}
%\end{figure}

In \FIG{p2}(b) we show $P_2(p,L)$ for SP and BP1 models. Here we also see the bond-site duality for BP1 and SP models.
Like $P_2$ in OBP, $p_{2max}$ for BP1 and SP models approaches to the estimated $p_c$ from \FIGS{ps}(a) and (b) as $L$ increases. 
However, in contrast to OBP, $P_2(p_{2max})$ for both models does not decrease and remains nearly at constant value as $L$ increases.
This behavior clearly shows that $P_2$
for both models does not vanish at $p \simeq p_c$ in the limit $L \rightarrow \infty$,
which provides a strong evidence for the discontinuous transition.

In \FIG{p2}(c) we display $P_2(p,L)$ for BS1 and SS models. As can be seen from \FIG{p2}(c), there also exists the bond-site duality for BS1 and SS models.
As shown in \FIG{p2}(c), $P_2$ for both BS1 and SS models manifests anomalous behavior. Unlike OBP,  $p_{2max}$ for both models hardly varies as $L$ increases and is very close to the estimated $p_c$ from \FIGS{ps}(c) and (d) regardless of $L$. 
Furthermore $P_2(p_{2max})$ for SS model remains  nearly at constant value or increases slightly as $L$ increases. 
These results for BS1 and SS models physically mean that there exists a stable macroscopic second largest cluster even at $p_c$ in the thermodynamic limit and 
the transition should be discontinuous.

In contrast, $P_2$ for BP2 and BS2 models in \FIG{p2}(d) is physically very similar to that for OBP in \FIG{p2}(a). $p_{2max}$ for BP2 and BS2 models approaches to the estimated $p_c$  as $L$ increases. $P_2(p_{2max})$ for BS2 model decreases as $L$ increases as that for OBP, which indicates that $P_2$ vanishes at $p \simeq p_c$ in the thermodynamic limit. 
Thus, the transition becomes continuous for both BP2 and BS2 as expected from $P_s(p)$ data.

%\section{summary and discussions}

In this letter we exactly define the explosive lattice percolation models on the square lattice.
By studying $P_s$ and $P_2$ for the models, we observed the bond-site duality in the pair of SP and BP1 models and in the pair of SS and BS1 models. The duality means the discontinuous transition. In contrast two bond models, BP2 and BS2 models, which relatively suppress the intra-bond occupation, undergo the continuous transition, which satisfies the normal scaling behavior like Eq. (1). 

%unneccesary references.
\begin{acknowledgments}
{This work was supported by National Research Foundation
of Korea (NRF) Grant funded by the Korean Government
(MEST) (Grant Nos. 2011-0015257)
and by Basic Science Research Program 
through the National Research Foundation of Korea(NRF) funded by
the Ministry of Education, Science and Technology (No. 2012R1A1A2007430).}
\end{acknowledgments}

\end{document}